\documentclass[prb,twocolumn,showpacs]{revtex4}

\usepackage{graphicx,color,ae,aecompl}
\usepackage[cm]{aeguill}
\usepackage{dcolumn}
\usepackage{bm}

\begin{document}
\title{Branching ratio measurement of a "Lambda" system in
  $\mathrm{Tm}^{3+}\mathrm{:YAG}$ under magnetic field}
\author{A. Louchet, J. S. Habib, V. Crozatier,  I. Lorger{\'e},
  F. Goldfarb, F. Bretenaker, J.-L. Le Gou\"et}
\affiliation{Laboratoire Aim\'e Cotton, CNRS-UPR 3321,  B\^atiment
505, Campus Universitaire, 91405 Orsay Cedex, France}
\author{O. Guillot-No\"el, Ph. Goldner}
\affiliation{ Laboratoire de Chimie de la Mati\`ere Condens\'ee de
Paris,
  CNRS-UMR 7574, ENSCP,\\ 11 rue Pierre et Marie Curie, 75231 Paris Cedex 05,
  France}
\date{\today}
\begin{abstract}
A three-level $\Lambda$ system in Tm${}^{3+}$ doped YAG crystal is
experimentally investigated in the prospect of quantum information
processing.  Zeeman effect is used to lift the nuclear spin
degeneracy of this ion. In a previous paper [de Seze \emph{et al.}
Phys. Rev. B, {\bf 73}, 85112 (2006)] we measured the gyromagnetic
tensor components and concluded that adequate magnetic field
orientation could optimize the optical connection of both ground
state sublevels to each one of the excited state sublevels, thus
generating $\Lambda$ systems. Here we report on the direct
measurement of the transition probability ratio along the two legs
of the Lambda. Measurement techniques combine frequency selective
optical pumping with optical nutation or photon echo processes.

\end{abstract}


 \maketitle

\section{Introduction}
\begin{figure}
\centering\includegraphics{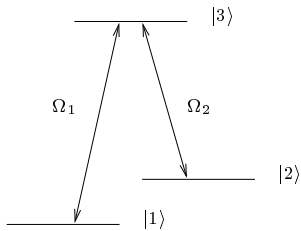} \caption{Three-level $\Lambda$
system coupled by two
  lasers. $\Omega_1$ and $\Omega_2$ are the Rabi frequencies that
  characterize the atom-laser interaction.}
\label{fig:lambda}
\end{figure}

There is an ongoing interest for macroscopic quantum processes,
especially in the perspective of quantum memories and several groups
are currently working on performing storage and retrieval of a
quantum state of light into an atomic system. Most quantum memory
processes rely on the storage of a quantum signal into an atomic
spin coherence, which is free from spontaneous emission. The optical
excitation carrying the information resonantly excites an atomic
transition. A control pulse then converts the optical coherence into
a hyperfine long-lived coherence. Another control pulse can
eventually change the spin coherence back into an optical coherence
and allow the retrieval of the stored signal. This quantum memory
scheme can be achieved in a three-level $\Lambda$ system, where two
spin states are connected to a common upper level by optical
transitions (cf Fig.~\ref{fig:lambda}).

The mapping of a quantum state of light over a macroscopic atomic
ensemble has been demonstrated in various materials such as atomic
vapors~\cite{julsgaard2001} or cold atom clouds~\cite{kuzmich2003}.
The full quantum memory scheme has recently been achieved in cold
atomic clouds of rubidium~\cite{chaneliere2005}. However, rare-earth
ion-doped crystals also appear as promising candidates for quantum
storage applications. They offer properties similar to atomic vapors
with the advantage of no atomic diffusion. At low temperature ($<4$
K), the optical coherence lifetime can reach several ms, and a
hyperfine coherence lifetime has been extended to tens of seconds in
Pr-doped $\mathrm{Y}_2\mathrm{SiO}_5$ (YSO)~\cite{fraval2004}. Given
the absence of atomic motion, extremely long population lifetime can
be observed. Electromagnetically induced transparency  and "slow
light"~\cite{turukhin2002} have been demonstrated in rare-earth
ion-doped crystals, along with storage times greater than a
second~\cite{longdell2005}.

Among all the rare-earth ions (lanthanides), non-Kramers ions with
an even number of $4f$ electrons exhibit the longest coherence
lifetime in the absence of external magnetic
field~\cite{macfarlane}. With the additional condition on the
hyperfine splitting to be of the order of a few tens of~MHz in the
electronic ground state, the rare earth ions are restricted to
praseodymium and europium. Both Pr${}^{3+}$ and Eu${}^{3+}$ ions
have been extensively studied over the past 20
years~\cite{mlynek1983,erickson1989,longdell2002,klieber2003,
babbitt1989,kroll1990,equall1994}. However, only dye lasers are
available at operation wavelengths in Pr${}^{3+}$ and Eu${}^{3+}$
doped crystals. For instance, $\mathrm{Pr}^{3+}\mathrm{:YSO}$ and
$\mathrm{Eu}^{3+}\mathrm{:YSO}$ are respectively operated at 606~nm
and 580~nm. Alternatively, other non-Kramers rare-earth ions such as
thulium fall within reach of more tractable lasers.

$\mathrm{Tm}^{3+}\mathrm{:Y}_3\mathrm{Al}_5\mathrm{O}_{12}$ (YAG)
has been widely studied in the field of coherent transient-based
signal-processing schemes~\cite{merkel1998,
lin1995,lavielle2002,gorju2005} and
lasers~\cite{scott1993,suni1991}. It is especially attractive
because the ${}^3H_6(0) \rightarrow {}^3H_4(0)$ transition at 793 nm
is in the range of semi-conductor lasers which can be stabilized to
better than 1 kHz~\cite{crozatier2004}.

Although the only isotope of thulium ${}^{169}$Tm ions has a $I=1/2$
nuclear spin, it does not exhibit hyperfine structure at zero
magnetic field because of $J$ quenching~\cite{abragam}. Lifting the
nuclear spin degeneracy with an external magnetic field offers a way
to build a $\Lambda$-system provided the magnetic field induces
different spin-state mixing in ground and excited electronic states.
In an earlier paper~\cite{deSeze2006} we showed that this was the
case for a specific magnetic field orientation with respect to the
crystalline axes. We determined the field direction that gives
maximum branching ratio between the two optical transition
probabilities in the $\Lambda$ system thus created. In addition, a
lower bound to the branching ratio was derived from the experimental
measurement of the gyromagnetic factors.

Further characterization of this $\Lambda$ system includes direct
experimental measurement of the optimal branching ratio. This is the
purpose of this study. The paper is arranged as follows. In
Sec.~\ref{lambda} we summarize the previous results concerning the
building of a $\Lambda$ system in $\mathrm{Tm}^{3+}\mathrm{:YAG}$.
In Sec.~\ref{exp} the experimental setup is described. In
Sec.~\ref{shb} spectral hole-burning is presented and discussed in
the case of $\mathrm{Tm}^{3+}\mathrm{:YAG}$, and we measure the
population lifetime of the ground sublevels. In Sec.~\ref{branching}
we show that each transition can be individually excited and we
measure their relative strengths by optical nutation and two-pulse
photon echoes. From these experiments we derive an experimental
value of the branching ratio.

\section{Building a $\Lambda$ system in
$\mathrm{Tm}^{3+}\mathrm{:YAG}$}\label{lambda} This section is a
brief reminder of the discussion presented in a previous
paper~\cite{deSeze2006}.

\begin{figure}
  \centering\includegraphics{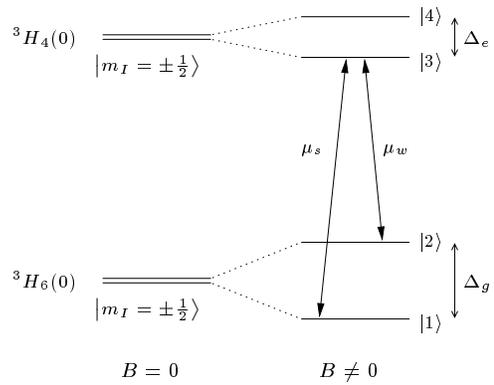} \caption{Building
a three-level system with nuclear spin levels in thulium.} \label{fig:rule}
\end{figure}

Let us consider the optical transition at 793~nm connecting the
fundamental levels of the Stark multiplets ${}^3H_6$ and ${}^3H_4$
of Tm${}^{3+}$ in YAG. Application of an external magnetic field
lifts the nuclear spin degeneracy by splitting the electronic
levels. As shown in Fig.~\ref{fig:rule}, a three-level $\Lambda$
system would involve the two hyperfine sublevels of the ground state
${}^3H_6(0)$ and one hyperfine sublevel of the excited state
${}^3H_4(0)$. If the interaction with the magnetic field is
restricted to the Zeeman effect, the electronic levels split into
$m_I=1/2$ and $m_I=-1/2$ spin levels. One of the two optical
transitions of the $\Lambda$ is forbidden. Indeed the selection rule
$\Delta m_I=0$ forbids any electronic excitation to flip the nuclear
spin. Fortunately, the coupling effect of the Zeeman electronic and
the hyperfine interactions mixes the nuclear spin states and may
lead to comparable transition strengths for the two legs of the
$\Lambda$ if the magnetic field is suitably oriented. As a
consequence, spin-flipping transitions
$\left|1\right>\rightarrow\left|4\right>$ and
$\left|2\right>\rightarrow\left|3\right>$ are no longer forbidden.
The atomic states can be represented as the product of an electronic
state and a nuclear spin state. The ground level nuclear spin states
are derived from the excited level nuclear spin states by a unitary
transformation. Therefore the 4 transition dipole moments satisfy:
\begin{equation}
\left\{ \begin{array}{ccccc}
  \left<1|\mu|3\right> &=& \left<2|\mu|4\right>&=&\mu_s\\
  \left<1|\mu|4\right> &=& - \left<2|\mu|3\right>&=&\mu_w
\end{array}
\right.
\end{equation}
The weak (respectively, strong) transition dipole moments are
referred to as $\mu_w$ (respectively, $\mu_s$). The branching ratio
$R$ is the ratio of the transition probabilities along the two legs
of the $\Lambda$:
\begin{equation}
R=\frac{\left| \left<2|\mu|3\right>\right|^2}{\left|
\left<1|\mu|3\right>\right|^2}= \frac{{\mu_w}^2}{{\mu_s}^2}
\end{equation}

\begin{figure}
\centering\includegraphics[width=8cm]{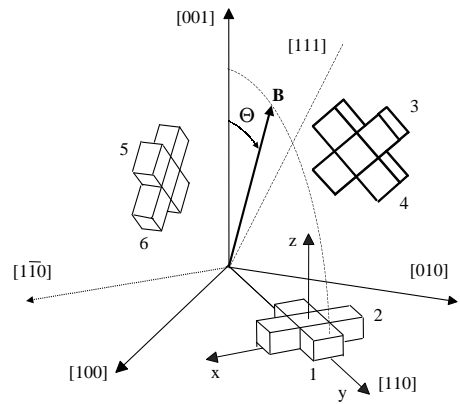} \caption{Tm
substitution sites in the YAG matrix. In each site, the transition
dipole moment $\vec{\mu}$ is represented by an oblong and is
directed along the local $Oy$ axis. Here the local frame $x,y,z$ is
given for site 1. The direction of the magnetic field ${\bf B}$ is
defined by angle $\Theta$.} \label{fig:sites}
\end{figure}

In the YAG crystal, $\mathrm{Tm}^{3+}$ ions substitute for yttrium
ions in 6 crystallographically equivalent but orientationally
inequivalent sites depicted in Fig.~\ref{fig:sites}. Because of the
site symmetry ($D_2$), the optical transition dipole moment is
oriented along the $Oy$ local axis of each site. Let the light beam
propagate along $[1\bar{1}0]$. Two linear polarization directions
are considered.

When it is directed along $[111]$, the optical electric field
$\bf{E}$ is perpendicular to the electric dipoles $\vec{\mu}$ of
sites 2, 4 and 6. The only excited sites are then 1, 3 and 5. They
interact with the light beam with the same Rabi frequency
$\Omega=\vec{\mu} \cdot {\bf E}/\hbar$ and therefore contribute the
same amount to the crystal optical density.

Alternatively, when the light beam is polarized along direction
$[\bar{1}\bar{1}1]$, the only excited sites are 1, 4 and 6. Their
Rabi frequencies are equal.

For a given magnetic field orientation, the field coordinates are
specific to each site local frame. Since the gyromagnetic tensor is
strongly anisotropic in both electronic levels, the 6 sites exhibit
different hyperfine splittings:
\begin{equation}
\Delta_{g,e}=\sqrt{ {\gamma_x^2}_{g,e} B_x^2+{\gamma_y^2}_{g,e}
B_y^2+{\gamma_z^2}_{g,e} B_z^2}
\end{equation}
where $g$ and $e$ denote the ground and excited electronic states,
and where the gyromagnetic tensor $\gamma$ and the magnetic field
${\bf B}$ coordinates are given in the local frame. Let ${\bf B}$ be
applied in the bisector plane defined by crystalline axes $[001]$
and $[110]$. The magnetic field direction is defined by angle
$\Theta$ formed with direction $[001]$, as illustrated in
Fig.~\ref{fig:sites}. For any magnetic field orientation within the
bisector plane, ions in sites 3 and 5 experience the same field.
Hence they have identical Zeeman splittings and equal branching
ratios. Similarly, ions in sites 4 and 6 experience the same
magnetic field and exhibit equal branching ratio.

\begin{figure}
\centering\includegraphics[width=8cm]{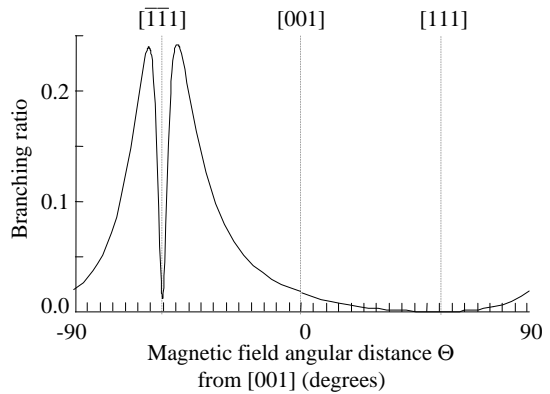} \caption{Theoretical
branching ratio variations for ions in sites 3 and 5 as a function
of magnetic field orientation~\cite{Guillot2005}.} \label{fig:br}
\end{figure}

The branching ratio for ions in sites 3 and 5 was calculated from
theoretical gyromagnetic factors for any magnetic field
orientation~\cite{Guillot2005}. When the magnetic field is in the
bisector plane, Fig.~\ref{fig:br} shows that the theoretically
predicted branching ratio goes through a maximum
$R_{\mathrm{max}}=0.24$. Besides, Fig.~\ref{fig:br} shows that the
accurate orientation of the magnetic field is critical around the
maximum branching ratio orientation.

The experimental measurement of the gyromagnetic tensor
coefficients~\cite{deSeze2006} made it possible to determine a lower
boundary for the maximum branching ratio:
\begin{equation}\label{eq:lower}
R_{\mathrm{max}}\ge 0.13\pm0.02
\end{equation}
These measurements also yielded the  magnetic field orientation that
gives maximum branching ratio for ions in sites 3 and 5:
$\Theta=-49.4\pm0.2^\circ$. This is rather close to direction
$[\bar{1}\bar{1}1]$ (corresponding to $\Theta=-\arccos
(1/\sqrt{3})=-54.8^\circ$).

\section{Experimental setup}\label{exp}
The system is excited and probed with an extended cavity
semiconductor laser operating at 793~nm. It is stabilized on a
high-finesse Fabry-Perot cavity through a Pound-Drever-Hall
servo-loop~\cite{crozatier2004}. It is amplified with a Toptica
BoosTA laser amplifier. A single-mode fiber spatially filters the
beam. At the end of the fiber one measures up to 100~mW output
power, depending on the amplifier adjustment.

The  $0.1$ at.\% $\mathrm{Tm}^{3+}\mathrm{:YAG}$ crystal is 5 mm
thick. It is cut perpendicular to direction $[1\bar{1}0]$ along
which the laser beam propagates (cf Fig.~\ref{fig:sites}). We use a
commercial Oxford 6T Spectromag SM4 cryostat. The adjustable
magnetic field is generated by superconducting coils immersed in
liquid helium. The sample is cooled down to $1.9$~K.

\begin{figure}
\centering\includegraphics[width=8cm]{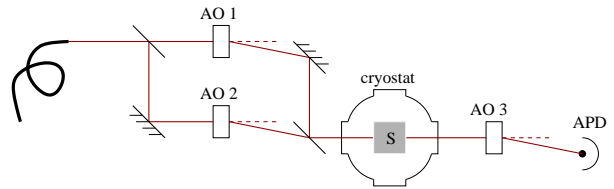} \caption{(color
online) Schematic of optical setup after the fiber. AO 1, 2 and 3
are acousto-optic devices, S is the Tm:YAG sample, and APD is the
avalanche photodiode. } \label{fig:setup}
\end{figure}

Fig.~\ref{fig:setup} shows the optical setup after the fiber. Pulse
sequences are shaped by an Arbitrary Waveform Generator (Tektronix
AWG 520) that monitors two external acousto-optic shifters (AO~1 and
AO~2). The acousto-optic shifters are imaged on the sample (S) so
that the beam does not move across the sample when the frequency is
varied. The spot size on the crystal is adjusted to 100~$\mu$m,
except in the experiments described in Sec.~\ref{sub}, where the
spot size is $800~\mu$m. The transmitted light is then collected on
an avalanche photodiode (APD) HAMAMATSU C5460. An acousto-optic
modulator (AO~3) protects the detector from possible strong burning
pulses.

\section{Hole burning spectroscopy}\label{shb}
\subsection{Spectral hole burning in an inhomogeneously broadened
4-level system}\label{hb} 
\begin{figure}
\centering\includegraphics{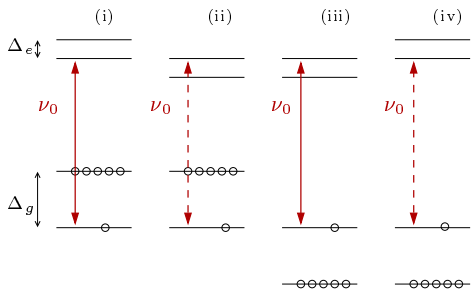} \caption{(color online)
  Optical pumping of four classes of ions to the non-resonant ground
  sublevel by excitation at frequency $\nu_0$. Solid lines: strong transitions. Dashed
  lines: weak transitions. Ions are depicted by open circles.
} \label{fig:4niv}
\end{figure}

In thulium, the optical transition at
793~nm exhibits a very large inhomogeneous broadening
$\Gamma_{\mathrm{inh}}\simeq 20$~GHz. Application of a magnetic
field splits each electronic level into a pair of hyperfine
sublevels with a Zeeman splitting of the order of $10$ to
$400$~MHz/T, depending on the magnetic field orientation with
respect to the local site axes. A field lower than 1~Tesla does not
modify the absorption spectrum since the inhomogeneous linewidth is
much larger than the Zeeman splitting. Because of inhomogeneous
broadening, a laser excitation at given frequency $\nu_0$
simultaneously excites 4 different classes of ions, as illustrated
in Fig.~\ref{fig:4niv}. Each class refers to a specific ground and
excited state sublevels that are connected by the incoming field.
Initially, in each ion class both ground sublevels are equally
populated. This is even true at working temperature ($T=2$~K), since
the level splitting is much smaller than $k_B T/h\simeq30$~GHz.

When the sample is exposed to sustained incident light at $\nu_0$,
it experiences optical pumping that tends to unbalance the ground
level state distribution. Ions in the resonantly excited sublevel
are transferred to the off-resonant sublevel (cf
Fig.~\ref{fig:4niv}). This gives rise to a hole in the absorption
spectrum at frequency $\nu_0$. Besides, since we consider 4-level
systems, additional features appear. Two side holes are observed at
$\nu_0\pm\Delta_e$, where the probe field excites transitions from
the depleted ground sublevel to the other excited state sublevel.
Conversely, transitions from an overpopulated ground sublevel to
either excited sublevel give rise to antiholes in the absorption
spectrum. Such antiholes are observed at $\nu_0\pm\Delta_g$ and
$\nu_0\pm(\Delta_g\pm\Delta_e)$.

Depending on the sublevels they connect, the various transitions
excited by the probe beam can be sorted as weak or strong ones. It
should be noted that at least one strong transition contributes to
any hole and antihole, except for the antiholes located at
$\nu_0\pm(\Delta_g+\Delta_e)$. These antiholes arise from a probing
process along a weak transition in the ion classes (ii) and (iv) of
Fig.~\ref{fig:4niv}. As a result, the observation of an antihole at
this position shall bear evidence that the nuclear spin flip
selection rule is actually relaxed.

\begin{figure}
\centering\includegraphics[width=8.6cm]{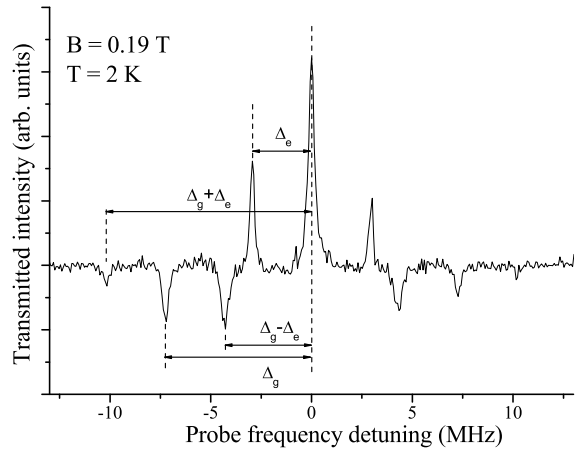}
\caption{Hole-burning transmission spectrum for
$\mathrm{Tm}^{3+}\mathrm{:YAG}$ under a $0.19$~T magnetic field. The
spectral holes and antiholes correspond to site 3 and 5 ions. The
features corresponding to site 1 are out of the probing spectral
window.} \label{fig:spectre}
\end{figure}

We perform spectral hole-burning in Tm${}^{3+}$:YAG. A 0.19~T
magnetic field is oriented in the direction that is expected to give
maximum branching ratio for ions in sites 3 and 5
($\Theta=-49.4^\circ$, cf Sec.~\ref{lambda}). The laser beam is
linearly polarized in direction $[111]$ so that only sites 1, 3 and
5 are excited. The burning step consists in a series of ten
$50~\mu$s pulses every 10~ms. Then the sample is probed with a weak
1 ms long pulse chirped over a $35$~MHz interval. The burning and
probing sequence is repeated every 150~ms. Transmitted probe
intensity is collected on a photodiode, averaged over 8 shots and
plotted in Fig.~\ref{fig:spectre}. The holes and antiholes  are
ascribed to sites 3 and 5. We measured the ground and excited
splittings for these two sites in this field direction:
\begin{eqnarray}
\Delta_g/B=38.2\pm 1~\mathrm{MHz/T} \\
\Delta_e/B=15.5 \pm 0.7~\mathrm{MHz/T}
\end{eqnarray}
where $B$ is the magnetic field amplitude. This agrees with the
calculated splittings derived from the experimental gyromagnetic
tensor components~\cite{deSeze2006}: $\Delta_g/B=36.0\pm 1$~MHz/T
and $\Delta_e/B=16.0 \pm 0.8$~MHz/T. The ground level gyromagnetic
tensor values are close to those measured by Schmidt in TmAG by
Nuclear Magnetic Resonance~\cite{schmidt1970}.

One can see in Fig.~\ref{fig:spectre} that the antiholes are
generally wider than the holes. This suggests that the inhomogeneous
broadening of the Zeeman splitting is larger in the ground state
than in the excited state.

The hole-burning features associated with site 1 cannot be seen in
Fig.~\ref{fig:spectre} because the Zeeman splitting for site 1 is
over 60 MHz/T for this magnetic field orientation.

\subsection{Sublevel population lifetime}
\label{sub} The hole-burning spectrum is erased by return to thermal
equilibrium. To measure the corresponding decay rate, we measure the
central hole depth as a function of delay between pumping and
probing steps. The sample is cooled down to $T= 1.9$~K. A $0.45$~T
magnetic field is applied in direction $\Theta=-54.5^\circ$, very
close to $[\bar{1}\bar{1}1]$. It should be noted that this is not
the optimal direction discussed in Sec.~\ref{lambda}.

\begin{figure}
\centering\includegraphics[width=8.6cm]{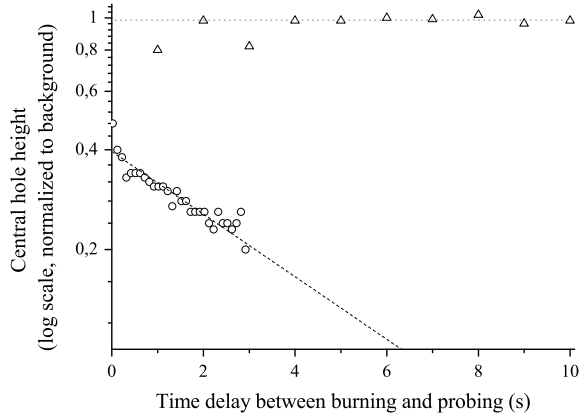} \caption{Decay of
central spectral hole after burning operation for sites 1, 4 and 6
(triangles) and sites 3 and 5 (circles). For these two measurements
an exponential function is fitted (dashed lines). In this experiment
only, the 0.45 T magnetic field is along $[\bar{1}\bar{1}1]$.}
\label{fig:pop}
\end{figure}

The light beam is polarized in direction $[\bar{1}\bar{1}1]$ so that
only sites 1, 4 and 6 are excited, with identical Rabi frequency. In
addition, since they are magnetically equivalent, their Zeeman
splittings are equal ($\Delta_g=181.4$~MHz and $\Delta_e=36.9$~MHz)
and they exhibit the same population lifetime. Fig.~\ref{fig:pop}
shows that the central hole does not decay for 10 seconds following
excitation. In other words, the population lifetime relative to
sites 1, 4 and 6 for this magnetic field orientation and strength is
estimated to a few minutes.

The light beam is then polarized along $[111]$ so as to excite sites
1, 3 and 5 with identical Rabi frequency. Since we are interested in
measuring the sublevel population lifetime for sites 3 and 5 only,
we isolate ions in sites 3 and 5 by means of a bleaching process: a
strong beam chirped over a 25~MHz interval optically pumps a large
spectrum range of ions in site 1 to the non-resonant very long
lifetime ground sublevel. Site 3 and 5 ions also undergo optical
pumping but the chirp amplitude is much larger than the ground
sublevel splitting $\Delta_g=6.9$~MHz. As a consequence, a large
fraction of ions in sites 3 and 5 is pumped back to the initial
sublevel. Three seconds after this bleaching process, we perform
ordinary spectral hole-burning and observe the decay of the central
hole depth by varying the delay between burning and probing (cf
Fig.~\ref{fig:pop}). An exponential fit gives a population lifetime
of $4.5 \pm 0.5$~s for sites 3 and~5.

Therefore in crystallographically equivalent sites 1,3 and 5, the
sublevel population lifetime apparently increases with the ground
state splitting. At present we cannot explain this behavior. In
contradiction with our measurements, N. Ohlsson \emph{et al.}
observed longer lifetimes for smaller splittings in a given
site~\cite{ohlsson2003}.

Since the spin level thermalization does not involve optical
transitions, we expect the corresponding decay rate to be
insensitive to a small angle rotation of the magnetic field. Indeed,
when ${\bf B}$ orientation is varied from $\Theta=-54.5^\circ$ to
$\Theta=-49.4^\circ$, the branching ratio of the optical transitions
strongly increases (cf Fig.~\ref{fig:br}), but the level splitting
$\Delta_g$ varies little, as shown in Table~\ref{tab:deltag}. The
decay rate is expected to be mainly sensitive to this level
splitting.

\begin{table}
\begin{tabular}{ccc}
\hline \hline
 & $\Theta=-54.5^\circ$  &  $\Theta=-49.4^\circ$
 (maximum $R$)\\
 \hline
site 1 & $328$ MHz/T & $306$ MHz/T\\
sites 3, 5 & $15$ MHz/T & $36$ MHz/T\\
\hline\hline
\end{tabular}
 \caption{Calculated values for normalized ground splitting $\Delta_g/B$
 in different sites, for two specific field orientations.}
 \label{tab:deltag}
\end{table}

Before sites 3 and 5 can be coherently manipulated for quantum
storage, it is necessary to prepare site 1 ions so that they do not
interfere. Their very long population lifetime guarantees that a
bleaching process can store them for a long time in a non-resonant
ground sublevel. As for sites 3 and 5, a spin level lifetime of a
few seconds is long enough for coherent manipulation.

\section{Branching ratio measurement}\label{branching}
In order to precisely compare the optical transitions along the two
legs of the $\Lambda$, we resort to coherent processes along with
frequency selective optical pumping. This enables us to isolate the
weak optical transition contribution.

\subsection{Optical nutation}\label{nutation}
A two-level system driven by a resonant monochromatic field ${\bf
E}$ oscillates between its ground and excited states at Rabi
frequency $\Omega=\vec{\mu}\cdot{\bf E}/\hbar$. Along with this
population oscillation, the system emits an induced radiation that
modulates the driving field at frequency $\Omega$. This gives a
direct and robust way to measure the Rabi frequency of a transition
even when the inhomogeneous broadening is infinite and the beam
spatial profile is not uniform: for an infinitely inhomogeneously
broadened medium driven by a stepfunction gaussian laser beam at
$t=0$, the transmitted signal takes the form of damped
oscillations~\cite{Sun2000}:
\begin{equation}
\label{eq:nut} \frac{I_{\mathrm{nut}}(t)}{I_0}=1-2(1-10^{-D})
\frac{J_1(\Omega t)}{\Omega t} \qquad \textrm{for } t> 0
\end{equation}
where $I_0$ is the incoming beam intensity, $D$ is the optical
density of the sample, $\Omega$ is the Rabi frequency of the
ion-laser interaction in the center of the beam and $J_1$ is the
Bessel function of order 1. This expression is valid if the optical
density of the sample is small ($D < 0.5$). According to
Eq.~\ref{eq:nut}, the first maximum of $I_{\mathrm{nut}}(t)$ is
located at $t_{\mathrm{max}}=5.1/\Omega$, therefore the Rabi
frequency can be directly obtained from the nutation signal.

The optical density $D$ can be assessed by comparing the transmission
of the material at $t=0$ with its transmission after a long exposure
to laser light:
\begin{equation}\label{eq:contrast}
D=\log_{10}\frac{I_{\mathrm{nut}}(t\rightarrow\infty)}{I_{\mathrm{nut}}(t=0)}
\end{equation}
This way, one can measure the optical density of the material with a
single nutation temporal profile. Besides, the optical density is
proportional to $\mu^2$ and to the number of active atoms. If
optical nutation is performed on each transition, then the optical
density measurement gives direct access to the relative dipole
moments of the weak and strong transitions.

\begin{figure}
\centering\includegraphics[width=8.6cm]{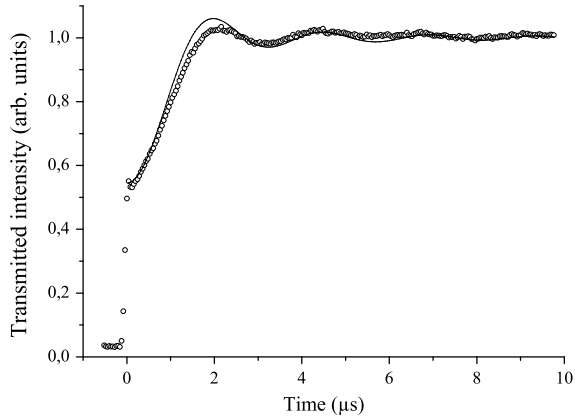} \caption{Optical
nutation signal with zero magnetic field (open circles) when a
step-function monochromatic excitation is applied at $t=0$. The
signal is fitted according to Eq.~\ref{eq:nut} with $\Omega=2.6$~MHz
and $D=0.32$ (solid line).} \label{fig:nut0}
\end{figure}

We first observe optical nutation on the two-level system formed by
$\mathrm{Tm}^{3+}\mathrm{:YAG}$ at zero magnetic field. The
transmitted signal is displayed in Fig.~\ref{fig:nut0}. The sample
is cooled down to $T=3$ K and is excited with a $10\ \mu$s resonant
light pulse polarized along $[111]$. For this light polarization,
sites 1, 3 and 5 participate in absorption with equal Rabi frequency
and each one of them contributes one third to the total optical
density $D_0=0.32$.

Let us apply a magnetic field in the direction that is expected to
give maximum branching ratio for sites 3 and 5 (cf
Sec.~\ref{lambda}). Application of a magnetic field splits the
electronic levels, the ions being equally distributed in both ground
sublevels. As mentioned in Sec.~\ref{hb}, this does not change the
absorption profile of the sample, so the nutation signal is not
modified.

We now apply $10~\mu$s pulses. The sample temperature is decreased
gradually from 5~K to 2~K. As long as thermalization processes
dominate, the system returns to thermal equilibrium between
successive pulses. As the temperature goes down, thermalization gets
slower. Ultimately, resonant sublevel depletion by optical pumping
supersedes repopulation by relaxation processes (see
Fig.~\ref{fig:4niv}) and the nutation signal disappears.

\begin{figure}
\centering\includegraphics{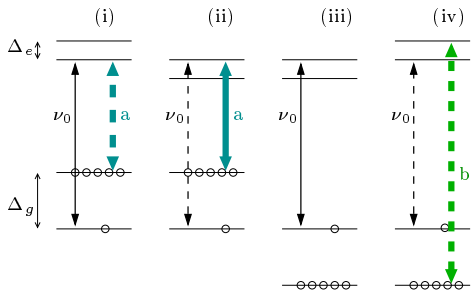} \caption{(color online)
  Examples of repumping schemes. Four classes of ions are excited at
  frequency $\nu_0$. The repumping frequencies are
  displayed as thick solid (respectively, dashed) arrows along strong
  (respectively, weak) transitions. (a) $\nu_R=\nu_0-\Delta_g$: only
  systems (i) and (ii) participate in the nutation signal, along a
  strong transition for system (i) and along a
  weak transition for system (ii).
  (b) $\nu_R=\nu_0+\Delta_g+\Delta_e$: only ions in system
  (iv) participate in the nutation signal, along a weak
  transition.}
\label{fig:repump}
\end{figure}

The repopulation process can be assisted by optical repumping. If
one tunes the repumping laser to $\nu_R=\nu_0-\Delta_g$ so as to
counterbalance optical pumping at $\nu_0$, only ions in sites 3 and
5 from classes (i) and (ii) are pumped back into the initial ground
sublevel [see Fig.~\ref{fig:repump}(a)]. All ions in site 1, as well
as ions in sites 3 and 5 with excitation schemes (iii) and (iv),
stay off resonance with the repumping beam. In a given class of
ions, the steady-state fraction of ions in the resonant sublevel
reads as
\begin{equation}
\rho=\frac{r+\kappa}{p+r+2\kappa}
\end{equation}
where $p$, $r$ and $\kappa$ respectively stand for the pumping rate,
the repumping rate and the temperature dependent relaxation rate. In
the absence of optical excitation this fraction reduces to
$\rho=1/2$, which means that the ground state population is evenly
shared between the two sublevels. When $r=0$ and $p\gg\kappa$, all
ions are pumped to the off-resonance sublevel and $\rho=0$. When
$r,p \gg\kappa$ the relative population reduces to $\rho=r/(p+r)$.

Optical pumping consists in repeated excitation and relaxation
processes. Excitation is carried out by a series of 10~$\mu$s
pulses. Within this 10~$\mu$s pulse duration the transition at
$\nu_0$ is bleached, resulting in evenly populated ground and upper
levels. As will be shown later in this section, this is true even
when the excitation takes place along the weak transition. Therefore
the excitation process brings the system in the same final state,
whatever excitation path is followed.

Relaxation to the ground state strongly differs from spontaneous
emission decay. Indeed, non-radiative processes dominate, proceeding
through the intermediate states ${}^3H_5$ and ${}^3F_4$. We assume
here that these stepwise relaxation processes relax the nuclear spin
selection rule. To sum up, neither excitation nor relaxation depends
on the transition strengths, and the same value of $\rho$ is used
for classes (i) and (ii).

The electronic transition dipole moment does not depart from its
zero magnetic field value. Since only 2 sites out of 3 participate
in the absorption, the optical density at stake here reads as
\begin{equation}
D_{s+w} =  \frac{2}{3}D_0\left(\frac{\mu_s^2 }{\mu_s^2+\mu_w^2} \
\rho+ \frac {\mu_w^2 }{\mu_s^2+\mu_w^2}\ \rho \right)= \frac{2}{3}
D_0\ \rho \label{eq:dsw}
\end{equation}
where $D_0$ is the optical density at zero magnetic field.

Similarly, if the repumping laser is tuned to
$\nu_R=\nu_0+\Delta_g+\Delta_e$ [cf Fig.~\ref{fig:repump}(b)], only
ions from system (iv) are repumped to the resonant sublevel: the
nutation signal is ascribed to ions in sites 3 and 5 resonating on a
weak transition. The optical density is expected to be much smaller
than $D_0$ and reads as
\begin{equation}\label{eq:dw}
D_w=\frac{2}{3}D_0  \frac{\mu_w^2 }{\mu_s^2+\mu_w^2}\
\rho=\frac{2}{3}D_0 \frac{R}{1+R}\ \rho
\end{equation}

Finally, if $\nu_R=\nu_0+\Delta_g-\Delta_e$, only  ions from system
(iii) are repumped to the resonant sublevel: the nutation signal is
ascribed to ions in sites 3 and 5 resonating on a strong transition.
The optical density reads as
\begin{equation}\label{eq:ds}
D_s=\frac{2}{3}D_0 \frac{\mu_s^2 }{\mu_s^2+\mu_w^2}\
\rho=\frac{2}{3}D_0 \frac{1}{1+R}\ \rho
\end{equation}

\begin{figure}
\centering\includegraphics[width=8.6cm]{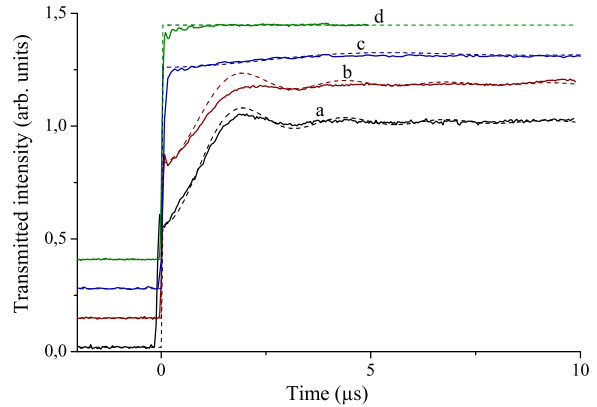} \caption{(color
online) Optical nutation signals for various repumping frequencies.
The four solid traces are obtained with the same incoming laser
intensity. a: with zero magnetic field, b: with magnetic field and
repumping at frequency $\nu_R=\nu_0-\Delta_g$, c: with magnetic
field and repumping at frequency $\nu_R=\nu_0-\Delta_g-\Delta_e$, d:
with magnetic field but no repumping beam. The dashed traces are
theoretical curves given by Eq.~\ref{eq:nut} and fitted to the
experimental data. The graphs are vertically offset for clarity.}
\label{fig:nut}
\end{figure}

The combination of optical pumping and repumping processes makes it
possible to isolate a nutation signal related either to a weak or to
a strong transition. The transmitted intensity for various repumping
frequency values is displayed in Fig.~\ref{fig:nut}. The repetition
rate is 50~s${}^{-1}$, the sample temperature is $T=2$~K and the
incident intensity is the same whatever the repumping frequency
value.

The optical density of the sample under zero magnetic field is
$D_0=0.36$ according to Eq.~\ref{eq:contrast} applied to
Fig.~\ref{fig:nut}(a). When a 0.3~T magnetic field is applied, the
sample becomes transparent: the damped oscillations disappear and
the transmitted intensity is almost constant [cf
Fig.~\ref{fig:nut}(d)].

The repumping process consists in a series of ten $100~\mu$s chirps
over 1~MHz around the repumping frequency $\nu_R$. It is applied
between successive excitation pulses. In Fig.~\ref{fig:nut}(b) we
plotted the transmitted intensity when a repumping beam is applied
at frequency $\nu_0-\Delta_g$. The optical density is measured on
the nutation signal: $D_{s+w}=0.167$. Eq.~\ref{eq:dsw} gives the
relative population of the resonant sublevel in systems (i) and
(ii): $\rho=69\%$. Besides, in both cases (a) and (b) of
Fig.~\ref{fig:nut}, the nutation signal is associated with exciting
strong transitions as well as weak transitions. This is why the Rabi
frequency and hence the position of the first maximum in the
nutation signal do not change.

Finally, in Fig.~\ref{fig:nut}(c) we plotted the transmitted
intensity for a repumping beam at $\nu_0+\Delta_g+\Delta_e$. The
optical density of the sample excited only along a weak transition
is very small: $D_w=0.018\pm0.004$. No damped oscillations are
visible, but the transmitted intensity slowly increases with time
during excitation. The experimental data clearly show that the
position of the first maximum in nutation signal (c) is different
from cases (a) and (b), even though it cannot be accurately located.
This proves that  signal (c) does not come from a residual signal
emitted along a strong transition.

According to Eqs.~\ref{eq:dsw} and~\ref{eq:dw}, the ratio of $D_w$
and $D_{s+w}$ finally leads to:
\begin{equation}
R= \frac{{\mu_w}^2}{{\mu_s}^2} =0.12\pm0.03.
\end{equation}
which is consistent with the lower boundary given in
Eq.~\ref{eq:lower} derived from gyromagnetic tensor
measurements~\cite{deSeze2006}.

The position of the first maximum of nutation profile for the weak
transition is $\sqrt{R}$ times the corresponding position for the
strong transition. According to the measured $R$ value, this maximum
occurs at $t\simeq5.5~\mu$s. Therefore, as previously assumed,
saturation is reached within the $10~\mu$s pulse duration even for
the weak transition.

\subsection{Photon echo}
In this section, two-pulse photon echoes are achieved separately on
each leg of the $\Lambda$. Application of a magnetic field splits
the electronic levels, and repeated excitation achieves optical
pumping between the two ground sublevels. Therefore the echo is
present as long as the temperature is high enough so that relaxation
compensates for optical pumping.

\begin{figure}
\centering\includegraphics[width=8.6cm]{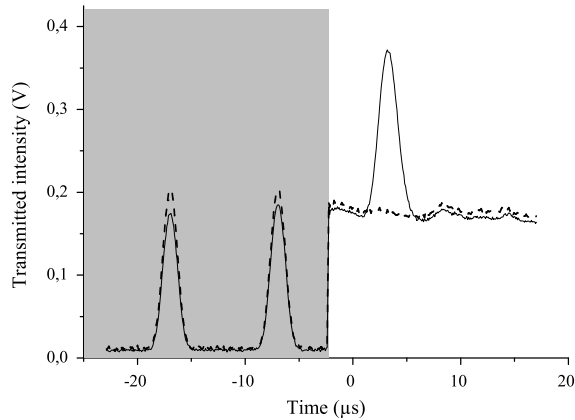} \caption{Photon
echo in $\mathrm{Tm}^{3+}\mathrm{:YAG}$ under magnetic field with
two gaussian pulses at $T=2.6$~K (solid line) and $T=1.9$~K (dashed
line). The detection channel is gated by an acousto-optic modulator
(AO~3 in Fig.~\ref{fig:setup}) which attenuates the excitation
pulses (grey area). This AOM opens a few $\mu$s after the second
pulse to let the photon echo through. The echo disappears when
optical pumping dominates thermalization.} \label{fig:echo}
\end{figure}

Fig.~\ref{fig:echo} shows a two-pulse photon echo sequence in Tm:YAG
under 0.3~T magnetic field, with a 50~s${}^{-1}$ repetition rate.
The two pulses are gaussian and have equal intensity. At $T=2.6$~K,
thermalization still dominates the optical pumping caused by
repeated excitation. In each two-pulse sequence, the first pulse
partly bleaches the optical transition, so the second pulse is less
absorbed than the first one. As the sample is cooled down, optical
pumping becomes dominant. At $T=1.9$~K, the resonant level is
totally empty and the photon echo signal disappears. The excitation
pulses propagate through a transparent medium where there is no
bleaching process, which is why their transmitted intensities are
equal.

We need to perform the same repumping process as described in
Sec.~\ref{nutation} in order to recover a signal and ascribe it to a
weak or strong transition. When the repumping frequency is tuned to
$\nu_0\pm(\Delta_g-\Delta_e)$, ions in system (i) or (iii) in
Fig.~\ref{fig:repump} are repumped and the photon echo is emitted
along the strong transition at frequency $\nu_0$. When the repumping
frequency is tuned to $\nu_0\pm(\Delta_g+\Delta_e)$, the photon echo
is emitted along the weak transition at frequency $\nu_0$ in ions in
system (ii) or (iv). To make sure that the detected signal is
imputable to the weak transition, we slightly detune the repumping
beam: the echo intensity goes down, which proves that the
off-resonant repumping is less efficient and that the signal does
not come from a residual signal emitted along a strong transition.

The photon echo amplitude can be expressed as a product of the
transition dipole moment and a function of the Rabi frequency. This
applies to both transitions, either strong or weak. Since the Rabi
frequency is proportional to the transition dipole moment and to the
applied field, the photon echo intensity along the strong transition
reads as:
\begin{equation}
\mathcal{I}_s (\mathcal{I})=\mu_s^2\ g(\mu_s^2 \mathcal{I})
\end{equation}
where $\mathcal{I}$ represents the laser intensity and where the
shape of function $g$ need not be detailed any further.

Conversely, when the weak transition is excited with the same
intensity $\mathcal{I}$, the photon echo intensity reads as
\begin{equation}
\mathcal{I}_w (\mathcal{I})=\mu_w^2 \ g(\mu_w^2 \mathcal{I}) = R \
\mu_s^2\ g(\mu_s^2 R\mathcal{I})=R \ \mathcal{I}_s(R \mathcal{I})
\end{equation}
where $R$ is the branching ratio.

\begin{figure}
\centering\includegraphics[width=8.3cm]{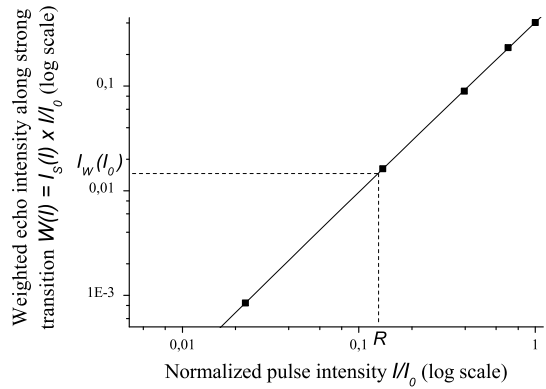} \caption{Weighted
echo intensity on the strong transition for different excitation
pulse intensities (squares). Theses values are interpolated with a
straight line (solid line) in order to solve Eq.~\ref{eq:implicit}.
A graphical solution is given by the intersection point of
$W(\mathcal{I})=\mathcal{I}_s( \mathcal{I})\times
\mathcal{I}/\mathcal{I}_0$ and
$W(\mathcal{I})=\mathcal{I}_w(\mathcal{I}_0)$. The value is directly
read out as $R=\mathcal{I}/\mathcal{I}_0$ at the intersection.}
\label{fig:Is}
\end{figure}

The photon echo intensity $\mathcal{I}_s$ on the strong transition
is measured for different excitation intensities. A weighted echo
intensity is defined as:
\begin{equation}\label{eq:W}
W(\mathcal{I})=\mathcal{I}_s( \mathcal{I})\times
\mathcal{I}/\mathcal{I}_0
\end{equation}
This function is interpolated with a polynomial function as shown in
Fig.~\ref{fig:Is}. If we measure $\mathcal{I}_w$ on the weak
transition for a given intensity $\mathcal{I}_0$, we are able to
derive the branching ratio value by solving the following equation:
\begin{equation}\label{eq:implicit}
\mathcal{I}_w(\mathcal{I}_0)=R \ \mathcal{I}_s(R \mathcal{I}_0)
\end{equation}
As shown in Fig.~\ref{fig:Is} and according to Eqs.~\ref{eq:W}
and~\ref{eq:implicit}, the graphical solution of this equation is
given by the intersection point of $W(\mathcal{I})=\mathcal{I}_s(
\mathcal{I})\times \mathcal{I}/\mathcal{I}_0$ and
$W(\mathcal{I})=\mathcal{I}_w(\mathcal{I}_0)$. The $x$-coordinate of
this intersection point is the branching ratio $R$:
\begin{equation}
R = 0.130 \pm 0.015.
\end{equation}
This is consistent too with the expectations derived from gyromagnetic
factor measurements~\cite{deSeze2006}.

\subsection{Discussion}
With optical nutation and photon echo experiments we measured two
branching ratio values that are in agreement with one another.

As far as theoretical gyromagnetic tensors are concerned, the
$(x,y)$ anisotropy is much larger in the ground state than in the
excited state~\cite{Guillot2005}. This difference, known as
\emph{anisotropy disparity}, is at the origin of the high value for
the maximum theoretical branching ratio $R_{\mathrm{max}}=0.24$. The
experimental value of anisotropy disparity appeared to be smaller
than predicted by theory~\cite{deSeze2006}. As a result, a branching
ratio value close to its lower boundary $0.13$ was expected.
The direct
measurements presented in this paper confirm this result.

These measurements  have been achieved for the magnetic field
orientation ($\Theta=49.4\pm0.2^\circ$) given by de Seze \emph{et
al.}~\cite{deSeze2006}. We did not study the dependence of $R$ as a
function of $\Theta$. However, the branching ratio is almost
constant over a $2^\circ$ wide interval~\cite{Guillot2005}.
Therefore, even with a $0.5^\circ$ mechanical precision on the
angle, one can reasonably consider that the measured branching ratio
is close to its maximum.

Finally, in the experiments presented in Sec.~\ref{nutation}, the
nutation signal along the strong transition goes through its first
maximum at $t=2~\mu$s. This means that a $\pi$-pulse along the
strong transition would take $1.2~\mu$s. According to the present
branching ratio measurements, a $\pi$-pulse along the weak
transition would then take $1.2/\sqrt{R}=3.4~\mu$s. Therefore, we
are able to excite each leg of the $\Lambda$ system with a
$\pi$-pulse, which is a capital asset in the perspective of coherent
processes dynamics.

\section{Conclusion}
We directly measure the transition probability ratio along the two
legs of the $\Lambda$ through coherent processes along with
frequency selective optical pumping. This is performed in the
optimum magnetic field orientation derived from a previous
experiment~\cite{deSeze2006}. The direct measurement of the
branching ratio is consistent with the value derived from the
gyromagnetic tensor components and confirms that a $\Lambda$-system
can operate efficiently in Tm:YAG. This way we have cross-checked
and supported the theoretical description, complementing prior
experimental data~\cite{deSeze2006} with a new set of data different
in nature. In addition, this provides us with the branching ratio
value we have to know in order to efficiently build and control
nuclear magnetic superposition states with optical excitation.

Side investigation on the sublevel lifetime leads to unexpected
results. Indeed, broadly spaced sublevels appear to be more stable
than closely spaced ones, in contradiction with previously reported
experiments~\cite{ohlsson2003}. This has to be clarified. Still, the
measured lifetime is consistent with long-time coherent manipulation
of spin states. The demonstrated existence of the $\Lambda$ system
promises interesting spin state properties in either ground or
excited level. Next step will deal with the investigation of these
coherent spin features.

\end{document}